\documentstyle[prl,aps]{revtex}
\twocolumn

\begin{document}
\author{Jian-Qi Shen$^{1,2}$\footnote{E-mail address: jqshen@coer.zju.edu.cn} and Zhi-Chao Ruan$^1$}
\address{$^1$ Centre for Optical
and Electromagnetic Research, Joint Research Centre of Photonics
of the Royal Institute of Technology (Sweden) and Zhejiang
University, Zhejiang University, Hangzhou Yuquan 310027, P. R.
China\\
$^2$ Zhejiang Institute of Modern Physics and Department of
Physics, Zhejiang University, \\Hangzhou 310027, P. R. China}
\date{\today }
\title{The time evolution of coherent atomic system and probe light
in an EIT medium} \maketitle

\begin{abstract}
The adiabatic solutions of Maxwell-Bloch equation governing the
three-level EIT medium is presented. The time evolution of the
density matrix elements of the EIT system and the probe light is
thus investigated by using the adiabatic approximation formulation
and the slowly varying envelope condition.


\end{abstract}

\section{Introduction}
Recently, many theoretical and experimental investigations show
that the control of phase coherence in multilevel atomic ensembles
will give rise to many novel and striking quantum optical
phenomena, such as the coherent population
trapping\cite{Arimondo}, laser without inversion and
electromagnetically induced transparency
(EIT)\cite{Zhusy,Zhuy2,Harris2}, in the wave propagation of
near-resonant light. According to the theoretical analysis of
multilevel atomic phase coherence, the requirement of the
occurrence of EIT is such that the strength of coupling light is
much stronger than that of probe light\cite{Harris2,Lukin}. Under
this condition, the EIT atomic vapor allows the probe light to
propagate without dissipation through the medium. Due to its
unusual quantum coherent character, the discovery of EIT has so
far led to many new peculiar effects and phenomena\cite{Harris2},
some of which are believed to be useful for the development of new
techniques in quantum optics. More recently, the physical effects
associated with EIT observed experimentally include the ultraslow
light pulse propagation\cite{Hau,Kash} and light
storage\cite{Liu,Phillips} in atomic vapor, and atomic ground
state cooling\cite{Morigi}.

With the development of the quantum information (quantum
computation, quantum communication and quantum measurement), the
search for new ways to manipulate photon states becomes
increasingly important\cite{Bouwmeester,Wheeler}. In the area of
EIT, historically, such a possibility might first be considered by
Ling {\it et al.}, who investigated the ``electromagnetically
induced grating'' in homogeneously broadened media\cite{Ling}. In
this work there is a strong coupling standing wave, interacting
with three-level Lambda-type (or cascade-type) atoms. This can
diffract a weak probe field, which propagates along a direction
normal to the standing wave, into high-order diffractions. As
stated by Arve {\it et al}., many researches of EIT including the
freezing light pulses\cite{Lukin2,Fleischhauer}, time-dependent
group velocity\cite{Fleischhauer2,Fleischhauer3}, delayed probe
light\cite{Liu}, stopping a pulse and restarting it in the
opposite direction (creating a time reversed version of the
original probe pulse)\cite{Zibrov} have been limited to
electromagnetic propagation in one dimension only\cite{Arve},
namely, the investigators of these studies did not consider how to
change the direction of the probe light. For this reason, Arve
{\it et al}. lifted the restriction to the one dimension and
considered the Maxwell-Bloch dynamics in two or more dimensions of
three-level EIT system, by using the adiabatic approximation and
the slowly varying envelope approximation. In their treatment,
however, they did not take into account the decay terms in the
dynamical equations. Since in a typical EIT
experiment\cite{Li,Schmidt}, the spontaneous decay rate of the
excited state of the three-level systems may be of the same order
of magnitude as that of the Rabi frequencies of the probe and
coupling optical fields, the contribution of the decay terms in
the Maxwell-Bloch dynamics should also be taken into
consideration. In the present paper, we obtain the adiabatic
solutions of the Maxwell-Bloch equations, {\it i.e.}, the
expressions for the time evolution of both the probe light and the
off-diagonal density matrix elements of the atomic system. The
formulation presented here can apply to the study of the variation
of the propagation direction (diffraction) of the optical fields
in the EIT medium.

\section{Maxwell-Bloch equations governing the three-level EIT system}
Consider a three-level $\Lambda$-type atomic ensemble interacting
with two resonant laser beams, the Rabi frequencies of which are
denoted by $\Omega_{\rm c}$ and $\Omega_{\rm p}$, respectively. In
such a $\Lambda$-type atomic system, levels $|c\rangle$ and
$|b\rangle$ are the ground states, and $|a\rangle$ the excited
state. The laser beam which couples levels $|a\rangle$ and
$|b\rangle$ is called the probe light ($\Omega_{\rm p}$). Another
laser beam is termed the coupling light ($\Omega_{\rm c}$), which
couples levels $|a\rangle$ and $|c\rangle$. The schematic diagram
for the above $\Lambda$-type system is depicted in Fig. 1.

If we apply the adiabatic approximation ({\it i.e.}, $|{{\rm
d}\rho}/{{\rm d}t}|\ll |\Omega_{\rm c}|$) to the three-level
system, then the Bloch equation can be rewritten in the form
\begin{equation}
\left\{
\begin{array}{ll}
& 0={\rm Im}\left(\Omega^{\ast}_{\rm p}\rho_{ab}+\Omega^{\ast}_{\rm c}\rho_{ac}\right)-\gamma_{aa}\rho_{aa},                 \\
& 0=\frac{i}{2}\left[\Omega_{\rm c}\rho_{cb}+\Omega_{\rm p}\left(\rho_{bb}-\rho_{aa}\right)\right]-\gamma_{ab}\rho_{ab},    \\
& 0=\frac{i}{2}\left[\Omega_{\rm p}\rho_{bc}+\Omega_{\rm c}\left(\rho_{cc}-\rho_{aa}\right)\right]-\gamma_{ac}\rho_{ac},     \\
& \dot{\rho}_{bb}={\rm Im}\left(\Omega_{\rm p}\rho_{ba}\right)-\gamma_{bb}\rho_{bb},       \\
& \dot{\rho}_{bc}=\frac{i}{2}\left(\Omega^{\ast}_{\rm p}\rho_{ac}-\Omega_{\rm c}\rho_{ba}\right)-\gamma_{bc}\rho_{bc},    \\
& 0={\rm Im}\left(\Omega_{\rm
c}\rho_{ca}\right)-\gamma_{cc}\rho_{cc},
\end{array}
\right. \label{eq1}
\end{equation}
where dot denotes the time derivative. In order to let level
$|b\rangle$ into a dark state (trapped state) and thus realize the
transparency effect for the probe light, the magnitude of Rabi
frequency $|\Omega_{\rm c}|$ should be much greater than $
|\Omega_{\rm p}|$. Generally speaking, $\Omega^{\ast}_{\rm
p}\rho_{ab}$ in the first equation in (\ref{eq1}) can be ignored
compared with the term $\Omega^{\ast}_{\rm c}\rho_{ac}$. It
follows from the first and last equations in (\ref{eq1}) that the
relation $\gamma_{aa}\rho_{aa}+\gamma_{cc}\rho_{cc}=0$ can be
satisfied, and consequently the ratio
 ${\rho_{aa}}/{\rho_{cc}}=-{\gamma_{cc}}/{\gamma_{aa}}$ is
obtained.

Assume that the initial state of the excited level $|a\rangle$ of
the EIT medium under consideration is nearly empty, and the two
ground states $|c\rangle$ and $|b\rangle$ are occupied according
to the dark state conditions, {\it i.e.}, $\rho_{bb}\simeq 1$ and
$\rho_{aa}\simeq 0$. This, therefore, means that the term
$\rho_{bb}-\rho_{aa}$ in the second equation in (\ref{eq1})
approximately equal unity. So, we can obtain the following
relation between the density matrix elements $\rho_{cb}$ and
$\rho_{ab}$
\begin{equation}
\rho_{cb}=-\frac{\Omega_{\rm
p}+2i\gamma_{ab}\rho_{ab}}{\Omega_{\rm c}}.    \label{eqeq1}
\end{equation}
In the meanwhile, it follows from the fifth equation in
(\ref{eq1}) that the relation between $\rho_{ba}$ and $\rho_{bc}$
is
\begin{equation}
\rho_{ba}=\frac{2i\left(\frac{\partial}{\partial
t}\rho_{bc}+\gamma_{bc}\rho_{bc}\right)}{\Omega_{\rm c}},
\label{eqeq2}
\end{equation}
where $\rho_{ba}$ and $\rho_{bc}$ are the complex conjugations to
$\rho_{ab}$ and $\rho_{cb}$, respectively. Note that Eqs.
(\ref{eqeq1}) and (\ref{eqeq2}) constitute a set of coupling
equations. In order to see how $\rho_{bc}$ and $\rho_{ba}$ evolve
in the presence of the probe and coupling lasers, we should first
obtain their respective equations. With the help of Eqs.
(\ref{eqeq1}) and (\ref{eqeq2}), one can arrive at
\begin{equation}
\frac{\partial}{\partial
t}\rho_{bc}+\left(\gamma_{bc}+\frac{\Omega^{\ast}_{\rm
c}\Omega_{\rm
c}}{4\gamma_{ab}}\right)\rho_{bc}+\frac{\Omega^{\ast}_{\rm
p}\Omega_{\rm c}}{4\gamma_{ab}}=0   \label{eqs4}
\end{equation}
and
\begin{equation}
\frac{\partial}{\partial
t}\rho_{ba}+\left(\gamma_{bc}+\frac{\Omega^{\ast}_{\rm
c}\Omega_{\rm
c}}{4\gamma_{ab}}\right)\rho_{ba}+\frac{i}{2\gamma_{ab}}\left(\frac{\partial}{\partial
t}\Omega^{\ast}_{\rm p}+\gamma_{bc}\Omega^{\ast}_{\rm p}\right)=0.
\label{eqrhoba}
\end{equation}
As in the adiabatic approximation, the Rabi frequency $\Omega_{\rm
c}$ of the coupling laser can be considered a constant in time,
one can derive
 $\rho_{bc}(t)$ from Eq. (\ref{eqs4}), {\it i.e.},
\begin{equation}
\rho_{bc}(t)=e^{-\lambda
t}\left[\int^{t}_{0}-\frac{\Omega^{\ast}_{\rm p}(t')\Omega_{\rm
c}}{4\gamma_{ab}}e^{\lambda t'}{\rm d}t'+\rho_{bc}(0)\right]
\label{eqbc}
\end{equation}
with $\lambda=\gamma_{bc}+{\Omega^{\ast}_{\rm c}\Omega_{\rm
c}}/({4\gamma_{ab}})$. In the meanwhile, one can obtain the
expression for $\rho_{ba}(t)$ from Eq. (\ref{eqrhoba})
\begin{eqnarray}
\rho_{ba}(t)&=& e^{-\lambda
t}\int^{t}_{0}\frac{1}{2i\gamma_{ab}}\left[\frac{\partial}{\partial
t'}\Omega^{\ast}_{\rm p}(t')+\gamma_{bc}\Omega^{\ast}_{\rm
p}(t')\right]e^{\lambda t'}{\rm d}t'      \nonumber   \\
& &+\rho_{ba}(0)e^{-\lambda t}.   \label{eqba}
\end{eqnarray}
It should be noted that although the Rabi frequency $\Omega_{\rm
c}$ is a time-dependent quantity, in the adiabatic case, it can be
viewed as a constant number. Such a result will be theoretically
validated by using the Maxwellian equations in the following.
Thus, the parameter $\lambda$ in Eqs. (\ref{eqbc}) and
(\ref{eqba}) is considered a constant number.

In the above, we consider the Bloch equation governing the
three-level EIT atomic media. In what follows, we will discuss the
Maxwellian equations
\begin{equation}
\left\{
\begin{array}{ll}
&  \left(c^{2}\nabla^{2}-\frac{\partial^{2}}{\partial
t^{2}}\right)\frac{1}{2} {\mathcal E}_{\rm c}e^{i({\bf k}_{\rm
c}\cdot {\bf r}-\omega_{\rm c}t)}
=\frac{Np_{ca}}{\epsilon_{0}}\frac{\partial^{2}}{\partial
t^{2}}\hat{\rho}_{ac},
 \\
&  \left(c^{2}\nabla^{2}-\frac{\partial^{2}}{\partial
t^{2}}\right)\frac{1}{2}{\mathcal E}_{\rm p}e^{i({\bf k}_{\rm
p}\cdot {\bf r}-\omega_{\rm
p}t)}=\frac{Np_{ba}}{\epsilon_{0}}\frac{\partial^{2}}{\partial
t^{2}}\hat{\rho}_{ab}
\end{array}
\right.     \label{Maxwell}
\end{equation}
of the coupling and probe lasers in an EIT medium, where $N$ and
$\epsilon_{0}$ denote the atomic number density (total number of
atoms per volume) and the electric permittivity in vacuum,
respectively. Note that here
 $\hat{\rho}_{ac}(t)$ and $\hat{\rho}_{ab}(t)$ can be rewritten as
 follows
$\hat{\rho}_{ac}(t)=\rho_{ac}(t)\exp\left[\frac{1}{i}\left(\omega_{ac}t-{\bf
k}_{ac}\cdot{\bf r}\right)\right]$,
$\hat{\rho}_{ab}(t)=\rho_{ab}(t)\exp\left[\frac{1}{i}\left(\omega_{ab}t-{\bf
k}_{ab}\cdot{\bf r}\right)\right]$. By using the slowly varying
envelope approximation ({\it i.e.}, the second order derivatives
of the envelopes ${\mathcal E}_{\rm c,p}$ are negligible compared
with the other terms on the left-handed sides of the above
Maxwellian equations), Eq. (\ref{Maxwell}) can be reduced to the
first order differential equation
\begin{equation}
\left\{
\begin{array}{ll}
& c\hat{{\bf k}}_{\rm c}\cdot\nabla{\mathcal E}_{\rm
c}+\frac{\partial}{\partial t}{\mathcal E}_{\rm c}
=i\frac{\omega_{\rm c}Np_{ca}}{\epsilon_{0}}\rho_{ac},               \\
& c\hat{{\bf k}}_{\rm p}\cdot\nabla{\mathcal E}_{\rm
p}+\frac{\partial}{\partial t}{\mathcal E}_{\rm
p}=i\frac{\omega_{\rm p}Np_{ba}}{\epsilon_{0}}\rho_{ab},
\end{array}
\right.
\end{equation}
where
\begin{eqnarray}
\rho_{ab}(t)&=& e^{-\lambda
t}\int^{t}_{0}\frac{1}{-2i\gamma_{ab}}\left[\frac{\partial}{\partial
t'}\Omega_{\rm p}(t')+\gamma_{bc}\Omega_{\rm
p}(t')\right]e^{\lambda t'}{\rm d}t'      \nonumber   \\
& &+\rho_{ab}(0)e^{-\lambda t}.
\end{eqnarray}
Here $p_{ca}$ and $p_{ba}$ stand for the transition dipole matrix
moments. Note that because of the initial condition ({\it e.g.},
the excited state $|a\rangle$ is nearly empty and, moreover, the
population probability of level $|c\rangle$ is negligibly small),
at least for the case of investigating the transient optical
properties of EIT media, $\rho_{ac}$ can be taken to be
zero\cite{Arve}. Thus the wave equation of the coupling light
${\mathcal E}_{\rm c}$ takes the form of a source-free field
equation. Since such an equation is simple, we will not further
consider it. Instead, in the next section we will concentrate our
attention on the wave equation of the probe light
\begin{equation}
c\hat{{\bf k}}_{\rm p}\cdot\nabla\Omega_{\rm
p}+\frac{\partial}{\partial t}\Omega_{\rm p}=i\frac{\omega_{\rm
p}Np_{ba}p_{ab}}{\epsilon_{0}\hbar}\rho_{ab},  \label{eqomega}
\end{equation}
where $\Omega_{\rm p}={{\bf {\mathcal E}_{\rm p}}\cdot{\bf
p}_{ab}}/{\hbar}$ with ${\bf p}_{ab}$ being the transition dipole
matrix moment of $ab$ transition.

It is apparently seen that if we have solved the solution
$\Omega_{\rm p}$ of Eq. (\ref{eqomega}), then according to
(\ref{eqbc}) and (\ref{eqba}), the information on the time
evolution of the medium polarization ($\rho_{bc}$ and $\rho_{ba}$)
can therefore be easily obtained. Such a study can apply to the
investigation of the adiabatic storing of a light pulse by the EIT
mechanism\cite{Arve}.

\section{Time evolution of the probe light}
The solution $\Omega_{\rm p}$ of Eq. (\ref{eqomega}) takes the
form
\begin{equation}
\Omega_{\rm p}({\bf r}, t)=f\left({\bf r}-\hat{{\bf k}}_{\rm
p}\int^{t}_{0}v_{\rm g}(t'){\rm d}t'\right),  \label{eq31}
\end{equation}
where $\hat{{\bf k}}_{\rm p}$ is a unit vector defined as
$\hat{{\bf k}}_{\rm p}={{{\bf k}}_{\rm p}}/{|{{\bf k}}_{\rm p}|}$,
and $v_{\rm g}(t)$ the group velocity of the probe light in the
EIT medium. Here $f$ is a certain function. It is readily verified
that the spatial derivative of $\Omega_{\rm p}$ can be expressed
in terms of the time derivative of $\Omega_{\rm p}$, {\it i.e.},
\begin{equation}
\hat{{\bf k}}_{\rm p}\cdot\nabla\Omega_{\rm p}=-\frac{1}{v_{\rm
g}}\frac{\partial}{\partial t}\Omega_{\rm p}. \label{eq3101}
\end{equation}
Thus, Eq. (\ref{eqomega}) can be rewritten as
\begin{equation}
\left(1-\frac{c}{v_{\rm g}}\right)\frac{\partial}{\partial
t}\Omega_{\rm p}=i\frac{\omega_{\rm
p}Np_{ba}p_{ab}}{\epsilon_{0}\hbar}\rho_{ab}.       \label{eq33}
\end{equation}
Further calculation ({\it i.e.}, calculating the time derivative
of the above equation) yields
\begin{equation}
\frac{\dot{v}_{\rm g}}{v^{2}_{\rm g}}\frac{\partial}{\partial
t}\Omega_{\rm p}+\left(1-\frac{c}{v_{\rm
g}}\right)\frac{\partial^{2}}{\partial t^{2}}\Omega_{\rm
p}=i\frac{\omega_{\rm
p}Np_{ba}p_{ab}}{\epsilon_{0}\hbar}\frac{\partial}{\partial
t}\rho_{ab}.       \label{eq34}
\end{equation}
In accordance with Eq. (\ref{eqrhoba}), we have
\begin{equation}
\frac{\partial}{\partial t}\rho_{ab}=-\lambda
\rho_{ab}+\frac{i}{2\gamma_{ab}}\left(\frac{\partial}{\partial
t}\Omega_{\rm p}+\gamma_{bc}\Omega_{\rm p}\right).    \label{eq35}
\end{equation}
Substitution of Eq. (\ref{eq33}) into (\ref{eq35}) yields
\begin{eqnarray}
\frac{\partial}{\partial t}\rho_{ab}&=&-\lambda
\left(i\frac{\omega_{\rm
p}Np_{ba}p_{ab}}{\epsilon_{0}\hbar}\right)^{-1}\left(1-\frac{c}{v_{\rm
g}}\right)\frac{\partial}{\partial t}\Omega_{\rm
p}            \nonumber \\
& &+\frac{i}{2\gamma_{ab}}\left(\frac{\partial}{\partial
t}\Omega_{\rm p}+\gamma_{bc}\Omega_{\rm p}\right).
 \label{eq36}
\end{eqnarray}
It follows from Eqs. (\ref{eq34}) and (\ref{eq36}) that the Rabi
frequency of the probe field, $\Omega_{\rm p}$, agrees with the
following equation
\begin{equation}
\frac{\partial^{2}}{\partial t^{2}}\Omega_{\rm p}+\zeta
\frac{\partial}{\partial t}\Omega_{\rm p}+\varsigma \Omega_{\rm
p}=0,                 \label{eq37}
\end{equation}
where the coefficients involved are defined as
\begin{equation}
\left\{
\begin{array}{ll}
& \zeta=\frac{1}{1-\frac{c}{v_{\rm g}}}\frac{\dot{v}_{\rm
g}}{v^{2}_{\rm g}}+\lambda+\frac{\omega_{\rm
p}Np_{ba}p_{ab}}{2\gamma_{ab}\epsilon_{0}\hbar\left(1-\frac{c}{v_{\rm
g}}\right)},                  \\
& \varsigma=\frac{\omega_{\rm
p}Np_{ba}p_{ab}\gamma_{bc}}{2\gamma_{ab}\epsilon_{0}\hbar\left(1-\frac{c}{v_{\rm
g}}\right)}.
\end{array}
\right.      \label{eq38}
\end{equation}
In this paper, we consider the homogeneous EIT media only, where
the group velocity of the probe light is independent of time, {\it
i.e.}, $\dot{v}_{\rm g}=0$. Thus the coefficients $\zeta$ and
$\varsigma$ in Eq. (\ref{eq37}) are constant. Such a choice will
simplify the problem under consideration, since the expression
(\ref{eq31}) can be reduced to the simple form $\Omega_{\rm
p}({\bf r}, t)=f\left({\bf r}-\hat{{\bf k}}_{\rm p}v_{\rm
g}t\right)$.

Apparently, the general solution of Eq. (\ref{eq37}) may be of the
form
\begin{equation}
\Omega_{\rm p}({\bf r}, t)=\tilde{\Omega}_{\rm p+}({\bf
r})\exp(\eta_{+}t)+\tilde{\Omega}_{\rm p-}({\bf
r})\exp(\eta_{-}t),     \label{eq39}
\end{equation}
where $\tilde{\Omega}_{\rm p\pm}({\bf r})$ is defined
\begin{equation}
\tilde{\Omega}_{\rm p\pm}({\bf r})=\Omega_{\rm p\pm}({\bf 0},
0)\exp\left(\sigma\hat{{\bf k}}_{\rm p}\cdot{\bf r}\right)
\label{eq310}
\end{equation}
with $\Omega_{\rm p\pm}({\bf 0}, 0)=\Omega_{\rm p\pm}({\bf r}={\bf
0}, t=0)$. Here $\sigma$ is a parameter that characterizes the
field amplitude shape. Keeping $\hat{{\bf k}}_{\rm
p}\cdot\hat{{\bf k}}_{\rm p}=1$ in mind, we rewrite the
time-dependent factor $\exp\left(\eta_{\pm}t\right)$ in Eq.
(\ref{eq39}) as follows
\begin{equation}
\exp\left(\eta_{\pm}t\right) \equiv  \exp\left[-\sigma \hat{{\bf
k}}_{\rm p}\cdot\hat{{\bf k}}_{\rm
p}\left(-\frac{\eta_{\pm}}{\sigma}\right)t\right].
\label{eq311}
\end{equation}
Hence Eq. (\ref{eq39}) is rewritten
\begin{eqnarray}
\Omega_{\rm p}({\bf r}, t)&=&\Omega_{\rm p+}({\bf 0},
0)\exp\left\{\sigma\hat{{\bf k}}_{\rm p}\cdot\left[{\bf r
}-\hat{{\bf k}}_{\rm
p}\left(-\frac{\eta_{+}}{\sigma}\right)t\right]\right\}        \nonumber \\
& &+\Omega_{\rm p-}({\bf 0}, 0)\exp\left\{\sigma\hat{{\bf k}}_{\rm
p}\cdot\left[{\bf r }-\hat{{\bf k}}_{\rm
p}\left(-\frac{\eta_{-}}{\sigma}\right)t\right]\right\}.
\label{eq312}
\end{eqnarray}
Comparing the expression (\ref{eq312}) with (\ref{eq31}), one can
arrive at the group velocity of the probe light
\begin{equation}
v_{\rm g}=-\frac{\eta_{\pm}}{\sigma},
\end{equation}
where $\eta_{\pm}=({-\zeta\pm\sqrt{\zeta^{2}-4\varsigma}})/{2}$,
which is expressed in terms of the parameters $\zeta$ and
$\varsigma$ of Eq. (\ref{eq37}). Note that here $v_{\rm g}$ is
constant, according to (\ref{eq38}), we have
\begin{equation}
\zeta=\lambda+\frac{\omega_{\rm
p}Np_{ba}p_{ab}}{2\gamma_{ab}\epsilon_{0}\hbar\left(1-\frac{c}{v_{\rm
g}}\right)}.
\end{equation}

Further calculation shows that the group velocity $v_{\rm g}$
satisfies the following quadratic equation
\begin{equation}
\sigma^{2}v^{2}_{\rm g}-\sigma v_{\rm g}\zeta+\varsigma=0.
\label{eq313}
\end{equation}
For convenience, we set $\zeta=\lambda+{\beta}({1-{c}/{v_{\rm
g}}})^{-1}$, $\varsigma={\beta\gamma_{bc}}({1-{c}/{v_{\rm
g}}})^{-1}$, where
\begin{equation}
\beta=\frac{\omega_{\rm
p}Np_{ba}p_{ab}}{2\gamma_{ab}\epsilon_{0}\hbar},
\end{equation}
and rewrite Eq. (\ref{eq313}) as
\begin{equation}
v^{2}_{\rm g}-\frac{\beta+\lambda+\sigma c}{\sigma}v_{\rm
g}+\frac{\lambda\sigma c+\beta\gamma_{bc}}{\sigma^{2}}=0,
\label{eq314}
\end{equation}
the two roots of which take the form
\begin{equation}
v_{\rm g\pm}=\frac{\beta+\lambda+\sigma c \pm
\sqrt{\left(\beta+\lambda+\sigma c\right)^{2}-4\left(\lambda\sigma
c+\beta\gamma_{bc}\right)}}{2\sigma}.
\end{equation}
Note that in a typical EIT experiment, the relationships between
the parameters $\gamma_{ab}$, $\gamma_{bc}$ and $\Omega_{\rm c}$
are as follows: $\gamma_{ab}\simeq \Omega_{\rm c}\simeq 10^{8}$
s$^{-1}$, $\gamma_{bc}\simeq 0.01\gamma_{ab}$ and $\omega_{\rm
p}\simeq 10^{15}$ s$^{-1}$\cite{Li,Schmidt}. Under these
conditions, one can verify that $\left(\beta+\lambda+\sigma
c\right)^{2}\gg 4\left(\lambda\sigma c+\beta\gamma_{bc}\right)$.
Thus, the group velocity, $v_{\rm g-}$, of the probe laser in the
EIT medium is reduced to
\begin{eqnarray}
v_{\rm g-}&=&\frac{\left(\beta+\lambda+\sigma
c\right)\left[1-\sqrt{1-\frac{4\left(\lambda\sigma
c+\beta\gamma_{bc}\right)}{\left(\beta+\lambda+\sigma
c\right)^{2}}}\right]}{2\sigma}
\nonumber \\
&\simeq& \frac{\lambda\sigma c+\beta\gamma_{bc}}{\sigma
\left(\beta+\lambda+\sigma c\right)}\rightarrow
\frac{\Omega^{\ast}_{\rm c}\Omega_{\rm c}}{\frac{2\omega_{\rm
p}Np_{ba}p_{ab}}{\epsilon_{0}\hbar}+\Omega^{\ast}_{\rm
c}\Omega_{\rm c}}c.      \label{vg}
\end{eqnarray}
It is worth pointing out that the above result (\ref{vg}) is
self-consistent, as it can be validated by using the formula of
group velocity, $v_{\rm g}=c/(n+\omega {\rm d}n/{\rm d}\omega)$,
where the optical refractive index
$n(\omega)={\sqrt{1+\chi(\omega)}}$. Here, the electric
susceptibility for the probe laser with a mode frequency $\omega$
in a steady EIT system is of the form \cite{Scully}
\begin{equation}
\chi(\omega)=-\frac{Np_{ba}p_{ab}}{\epsilon_{0}\hbar}\frac{\omega-\omega_{ab}+i\gamma_{bc}}{\left(\omega-\omega_{ab}+i\gamma_{bc}\right)\left(\omega-\omega_{ab}+i\gamma_{ab}\right)-\frac{\Omega^{\ast}_{\rm
c}\Omega_{\rm c}}{4}}.      \label{chi}
\end{equation}

In addition, it should be noted that the group velocity $v_{\rm
g+}$ ($v_{\rm g+}=-\eta_{-}/\sigma$) should be ruled out, since it
does not satisfy the relation
$\eta_{-}=({-\zeta-\sqrt{\zeta^{2}-4\varsigma}})/{2}$. So, the
only retained group velocity of the probe laser in the EIT medium
is $v_{\rm g-}$.

Eqs. (\ref{eq31}) and (\ref{eq3101}) shows that the adiabatic
solutions obtained in this paper can deals with the problems of
two- or three-dimensional wave propagation, including diffraction.
As is shown in Eq. (\ref{vg}), the coherent control of the probe
light can be realized: specifically, the change of $\Omega_{\rm
c}$ will has an influence on the probe group velocity (and hence
the parameters $\zeta$ and $\varsigma$). Thus, according to Eqs.
(\ref{eq31}), (\ref{eq3101}) and (\ref{eq37}), the variation of
the direction of propagation of the probe light may arise in
response to the change of the field strength of the coupling
laser.

\section{Off-diagonal density matrix elements}
In the previous section, we considered the wave propagation of the
probe laser. In this section, we will discuss the time evolution
of the off-diagonal density matrix elements of the EIT atomic
system. By inserting the expression (\ref{eq39}) into Eqs.
(\ref{eqbc}) and (\ref{eqba}), one can obtain the explicit
expressions for $\rho_{bc}$ and $\rho_{ba}$ in the time-evolution
process, {\it i.e.},
\begin{eqnarray}
\rho_{bc}(t)&=&-\frac{\Omega_{\rm
c}^{\ast}}{4\gamma_{ab}}[(\eta_{+}+\lambda)^{-1}\tilde{\Omega}^{\ast}_{\rm
p+}(e^{\eta_{+}t}-e^{-\lambda t})
                 \nonumber \\
& &+ (\eta_{-}+\lambda)^{-1}\tilde{\Omega}^{\ast}_{\rm
p-}(e^{\eta_{-}t}-e^{-\lambda t})]+\rho_{bc}(0)e^{-\lambda t},
 \nonumber \\
\rho_{ba}(t)&=&\frac{1}{2i\gamma_{ab}}[(\eta_{+}+\gamma_{bc})(\eta_{+}+\lambda)^{-1}\tilde{\Omega}^{\ast}_{\rm
p+}(e^{\eta_{+}t}-e^{-\lambda t})                 \nonumber \\
& &+
(\eta_{-}+\gamma_{bc})(\eta_{-}+\lambda)^{-1}\tilde{\Omega}^{\ast}_{\rm
p-}(e^{\eta_{-}t}-e^{-\lambda t})]
                            \nonumber \\
& &+\rho_{ba}(0)e^{-\lambda t}.    \label{rhoab}
\end{eqnarray}
In a typical EIT experiment ($N=10^{18}$ m$^{-3}$\cite{Schmidt}),
one can have $\zeta\gg \varsigma$. So, $\eta_{+}$, {\it i.e.},
$({-\zeta+\sqrt{\zeta^{2}-4\varsigma}})/{2}$ will be reduced to
$-{\varsigma}/{\zeta}$ that is less than $\gamma_{bc}$. In the
meanwhile, $\lambda=\gamma_{bc}+{\Omega^{\ast}_{\rm c}\Omega_{\rm
c}}/({4\gamma_{ab}})$, which is two or three orders of magnitude
larger than $\gamma_{bc}$ in a typical EIT experiment
\cite{Li,Schmidt}. This, therefore, means that when the time $t$
is taken the several Rabi oscillation periods ($\sim 10^{-8}$ s),
$e^{\eta_{+}t}\simeq 1 $ and $e^{-\lambda t}\simeq 0$. Thus,
$\rho_{ba}$ in (\ref{rhoab}) approximately equals
\begin{equation}
\rho_{ba}\simeq
\frac{1}{2i\gamma_{ab}}\frac{\gamma_{bc}}{\lambda}\tilde{\Omega}^{\ast}_{\rm
p+}
\end{equation}
or
\begin{equation}
\rho_{ba}\simeq
-\frac{i}{2}\frac{\gamma_{bc}}{\gamma_{ab}\gamma_{bc}+\frac{\Omega^{\ast}_{\rm
c}\Omega_{\rm c}}{4}}\tilde{\Omega}^{\ast}_{\rm p+}.
\end{equation}

According to Eq. (\ref{eq39}), we have $\Omega_{\rm p}({\bf r},
t)\simeq \tilde{\Omega}_{\rm p+}({\bf r})$, $\Omega_{\rm p}({\bf
r}, t)^{\ast}\simeq \tilde{\Omega}_{\rm p+}^{\ast}({\bf r})$.
Thus, by using the definition of the dipole-transition electric
susceptibility
\begin{equation}
\chi=\frac{2N|p_{ab}|^{2}}{\epsilon_{0}\hbar\tilde{\Omega}_{\rm
p+}}\rho_{ab},
\end{equation}
one can arrive at the electric susceptibility at probe frequency,
{\it i.e.},
\begin{equation}
\chi(\omega_{\rm
p})=\frac{N|p_{ab}|^{2}}{\epsilon_{0}\hbar}\frac{i\gamma_{bc}}{\gamma_{ab}\gamma_{bc}+\frac{\Omega^{\ast}_{\rm
c}\Omega_{\rm c}}{4}}.
\end{equation}
It should be noted that this expression is consistent with
(\ref{chi}), so long as the relation $\omega=\omega_{ab}$ (the
resonance condition) is inserted into (\ref{chi}). It is thus
believed that the obtained adiabatic solutions (\ref{rhoab}) of
Maxwell-Bloch equation governing the three-level EIT medium may be
self-consistent. This set of solutions can be utilized to consider
the information storage in a atomic vapor.

\section{Magnetic-dipole transition and permeability for the probe light}
In the previous sections, when treating the propagation of the
optical fields in the three-level system, we did not consider the
magnetic-dipole transition, but the electric-dipole transition. It
is believed that the contribution of the magnetic-dipole
transition to the wave propagation of the probe light should be
taken into account in the case of a stronger coupling field in the
EIT medium. According to the appendix to this paper, the density
matrix elements ${\rho}_{ab}$ and ${\rho}_{cb}$ satisfy the
following matrix equation
\begin{equation}
\frac{\partial}{\partial t}\left( {\begin{array}{*{20}c}
   {{\rho}_{ab}}  \\
   {{\rho}_{cb}}  \\
\end{array}} \right)\simeq
\left( {\begin{array}{*{20}c}
   {-\Gamma_{ab}} & {\frac{i}{2}\Omega_{\rm c}}  \\
   {\frac{i}{2}\Omega_{\rm c}^{\ast}} & {-\Gamma_{bc}}  \\
\end{array}} \right)\left( {\begin{array}{*{20}c}
    {{\rho}_{ab}}  \\
   {{\rho}_{cb}}  \\
\end{array}} \right)+\left( {\begin{array}{*{20}c}
    {\frac{i}{2}\Omega_{\rm p}}  \\
   {0}  \\
\end{array}} \right)
\label{density}.
\end{equation}
Here, $\Gamma_{ab}=\gamma_{ab}+i\Delta_{ab}$ and
$\Gamma_{bc}=\gamma_{bc}+i(\Delta_{ab}-\Delta_{ac})$, where
$\gamma_{ab}$ and $\gamma_{bc}$ represent the spontaneous decay
rate of level $|a\rangle$ and the dephasing rate (nonradiative
decay rate) of $|c\rangle$, respectively. It can be readily
verified that the steady solution of Eq. (\ref{density}) takes the
following form
\begin{equation}
\left\{
\begin{array}{ll}
&    {\rho}_{ab}=\frac{i\Omega_{\rm
p}\left[\gamma_{bc}+i(\Delta_{ab}-\Delta_{ac})\right]}{2\left\{\left(\gamma_{ab}+i\Delta_{ab}\right)\left[\gamma_{bc}+i(\Delta_{ab}-\Delta_{ac})\right]+\frac{\Omega_{\rm
c}^{\ast}\Omega_{\rm c}}{4}\right\}},      \\
&  {\rho}_{cb}=-\frac{\Omega_{\rm p}\Omega_{\rm c}^{\ast}}{4
\left\{\left(\gamma_{ab}+i\Delta_{ab}\right)\left[\gamma_{bc}+i(\Delta_{ab}-\Delta_{ac})\right]+\frac{\Omega_{\rm
c}^{\ast}\Omega_{\rm c}}{4}\right\}}.
\end{array}
\right. \label{steadysolution}
\end{equation}
Apparently, there exists a relation between $\tilde{\rho}_{cb}$
and $\tilde{\rho}_{ab}$, {\it i.e.},
\begin{equation}
{\rho}_{cb}=\frac{i}{2}\left\{\frac{\Omega_{\rm
c}^{\ast}}{\gamma_{bc}+i(\Delta_{ab}-\Delta_{ac})}\right\}{\rho}_{ab}.
\label{relationshipss}
\end{equation}
Note that since the level pairs $|a\rangle$-$|c\rangle$ and
$|a\rangle$-$|b\rangle$ can be coupled to two laser fields, the
parity of level $|a\rangle$ is different from both $|b\rangle$ and
$|c\rangle$. If, for example, $|a\rangle$ possess an odd parity,
$|b\rangle$ and $|c\rangle$ will have an even parity. Thus, the
electric dipole matrix elements ${\bf p}_{ab}=\langle a|e{\bf
r}|b\rangle\neq 0$ and ${\bf p}_{cb}=\langle c|e{\bf
r}|b\rangle=0$, and the magnetic dipole matrix elements ${\bf
m}_{cb}=\langle c|(e/2m_{\rm e})({\bf L}+2{\bf S})|b\rangle \neq
0$ and ${\bf m}_{ab}=\langle a|(e/2m_{\rm e})({\bf L}+2{\bf
S})|b\rangle=0$, where ${\bf L}$ and ${\bf S}$ denote the
operators of the orbital angular momentum and spin of electrons,
respectively. So, it is possible for the nearly resonant probe
laser to cause the electric-dipole transition between levels
$|a\rangle$ and $|b\rangle$ as well as the magnetic-dipole
transition between levels $|c\rangle$ and $|b\rangle$ in the
three-level atomic medium. The electric-dipole transition
($|a\rangle$-$|b\rangle$) and the magnetic-dipole transition
($|c\rangle$-$|b\rangle$) will yield the electric polarizability
and the magnetic susceptibility at probe frequency, respectively.
In general, the dimensionless ratio $|{\bf m}_{cb}/\left({\bf
p}_{ab}c\right)|\simeq 10^{-2}$ in an atomic system, where $c$
denotes the speed of light in a vacuum. For this reason, the
magnetic-dipole transition may not be considered in the treatment
for the wave propagation in an artificially electromagnetic
material. However, in a three-level EIT medium where the intensity
of coupling laser is much larger than that of the probe light, the
population in level $|c\rangle$ is much greater than that in the
upper level $|a\rangle$. In other words, the stronger coupling
laser enhances the probability amplitude of level $|c\rangle$
(according to the relation (\ref{relationshipss}), for the case of
small detunings, $|\rho_{cb}|\sim |(\Omega_{\rm
c}/\gamma_{bc})\rho_{ab}|$. In a typical EIT experiment
\cite{Li,Schmidt}, $\Omega_{\rm c}/\gamma_{bc}\simeq 10\sim 100$).
Thus, the order of magnitude of the density matrix element
$\rho_{cb}$ may be larger than that of $\rho_{ab}$. This,
therefore, means that the magnetic dipole moment
($2{m}_{cb}^{\ast}\rho_{cb}$) may possibly have the same order of
magnitude as the electric dipole moment
($2c{p}_{ab}^{\ast}\rho_{ab}$). As to the problem of local field
correction, here, in the simplest case (valid for gases), we can
take the local field to be the same as the macroscopic field
(average field in the sample). So, we need not consider the
Clausius-Mossotti-Lorentz relation. In a three-level atomic
system, the electric polarizability $\chi_{\rm e}$ and the
magnetic susceptibility $\chi_{\rm m}$ are of the form $\chi_{\rm
e}=2Np_{ab}^{\ast}{\rho}_{ab}/\left(\epsilon_{0}{\mathcal
E}\right)$ and $\chi_{\rm m}=2Nm_{cb}^{\ast}{\rho}_{cb}/{\mathcal
H}$, respectively, where $N,{\mathcal E}$ and ${\mathcal H}$
denote the atomic density (total number of atoms per volume), the
electric and magnetic field envelopes, respectively. Thus, with
the help of the steady solution (\ref{steadysolution}), one can
obtain the relative permittivity $\epsilon_{\rm r}=1+\chi_{\rm e}$
and the relative permeability $\mu_{\rm r}=1+\chi_{\rm m}$ of the
above atomic system. By using the relation ${\mathcal
H}=\sqrt{\epsilon_{\rm r}\epsilon_{0}/\mu_{\rm r}\mu_{0}}{\mathcal
E}$ between the envelopes of magnetic and electric fields, we have
the expressions for the electric polarizability and the magnetic
susceptibility, {\it i.e.},
\begin{equation}
\left\{
\begin{array}{ll}
& \chi_{\rm
e}=\frac{iN|p_{ab}|^{2}\left[\gamma_{bc}+i(\Delta_{ab}-\Delta_{ac})\right]}{\epsilon_{0}\hbar\left\{\left(\gamma_{ab}+i\Delta_{ab}\right)\left[\gamma_{bc}+i(\Delta_{ab}-\Delta_{ac})\right]+\frac{\Omega_{\rm
c}^{\ast}\Omega_{\rm c}}{4}\right\}},     \\
&   \chi_{\rm
m}=\frac{m_{cb}^{\ast}}{p_{ab}^{\ast}c}\sqrt{\frac{1+\chi_{\rm
m}}{1+\chi_{\rm e}}}\left[\frac{i}{2}\frac{\Omega_{\rm
c}^{\ast}}{\gamma_{bc}+i(\Delta_{ab}-\Delta_{ac})}\right]\chi_{\rm
e}.
\end{array}
\right. \label{expressions}
\end{equation}
It follows that the larger is the Rabi frequency $\Omega_{\rm c}$,
the more significant is the magnetic-dipole transition caused by
the quantum interference. So, the magnetic-dipole transition
deserves consideration in treating the wave propagation of the
probe light in the case of stronger coupling field.

It is shown that such an EIT medium may have the negative
permittivity and permeability, and will therefore become an ideal
candidate for realizing isotropic left-handed media, the optical
refractive index of which for the probe light is a negative
number. Such peculiar media exhibit a number of novel
electromagnetic and optical properties, including reversals of
both the Doppler shift and Cherenkov radiation, anomalous
refraction and amplification of evanescent
wave\cite{Veselago,Smith,Shelby,Shen}. Since this subject (the
scheme of EIT-based realization of negative refractive index
materials) is beyond the scope of the present paper, here we will
not consider it further, but will publish it elsewhere.
\section{Concluding remarks}

We treat the wave propagation of the probe light in a three-level
EIT medium and then obtain the solutions of Maxwell-Bloch
equations (containing the decay terms) by using the adiabatic
approximation under the slowly varying envelope condition. The
expressions for the time evolution of both the probe light and the
off-diagonal density matrix elements are presented, which can be
used to consider the variation of the propagation vector of the
probe light. This property, {\it i.e.}, the change of the
propagation direction and polarization of light in an EIT medium
by using coherent control may be important for the controllable
manipulation of optical fields ({\it e.g.}, light storage), as
well as for the quantum information processing. Under the
conditions of parameters which are taken as in the ordinary EIT
experiments, the results ({\it e.g.}, the group velocity and the
electric susceptibility for the probe light) obtained here can be
reduced to the ones derived from the steady solutions of Bloch
equations\cite{Scully}.

\textbf{Acknowledgements}  This work is supported by the National
Natural Science Foundation of China under Project Nos. $90101024$
and $60378037$.

\textbf{Appendix}

The three-level density matrix satisfies the following equation
\begin{equation}
\frac{\partial}{\partial t}\hat{\rho}=-i\left[\hat{H},
\hat{\rho}\right]-\frac{1}{2}\left\{\Gamma, \hat{\rho}\right\}
\eqnum{A1}
 \label{A1}
\end{equation}
with the Hamiltonian
\begin{equation}
\hat{H}= \left( {\begin{array}{*{20}c}
   {\omega_{a}} & {-\frac{1}{2}\Omega_{ab}} & {-\frac{1}{2}\Omega_{ac}} \\
  {-\frac{1}{2}\Omega_{ba}} & {\omega_{b}} & {0}  \\
    {-\frac{1}{2}\Omega_{ca}} & {0} & {\omega_{c}}  \\
\end{array}} \right),
\eqnum{A2} \label{eqH}
\end{equation}
where $\Omega_{ab}$, $\Omega_{ac}$ denote the Rabi frequencies of
two optical fields (the probe and coupling lasers) coupled to the
$ab$ and $ac$ transitions, respectively. $\Gamma$ is a diagonal
decay matrix of the levels, which agrees with $\langle i|\Gamma
|j\rangle=\gamma_{ii}\delta_{ij}$. $\omega_{a}$, $\omega_{b}$ and
$\omega_{c}$ are the level frequencies of $|a\rangle$, $|b\rangle$
and $|c\rangle$, respectively. If the density matrix can be
rewritten as
$\hat{\rho}(t)=\rho(t)\exp\left[\frac{1}{i}\left(\omega_{ij}t-{\bf
k}_{ij}\cdot{\bf r}\right)\right]$, then it follows from Eq.
(\ref{eqH}) that
\begin{equation} \left\{
\begin{array}{ll}
&  \dot{\rho}_{aa}={\rm Im}\left(\Omega^{\ast}_{\rm p}\rho_{ab}
+\Omega^{\ast}_{\rm c}\rho_{ac}\right)-\gamma_{aa}\rho_{aa},                  \\
& \dot{\rho}_{ab}=-i\Delta_{ab}\rho_{ab}
+\frac{i}{2}\left[\Omega_{\rm c}\rho_{cb}+\Omega_{\rm
p}\left(\rho_{bb}
-\rho_{aa}\right)\right]-\gamma_{ab}\rho_{ab},   \\
& \dot{\rho}_{ac}=-i\Delta_{ac}\rho_{ac}
+\frac{i}{2}\left[\Omega_{\rm p}\rho_{bc}+\Omega_{\rm
c}\left(\rho_{cc}-\rho_{aa}\right)\right]
-\gamma_{ac}\rho_{ac},    \\
& \dot{\rho}_{bb}={\rm Im}\left(\Omega_{\rm p}\rho_{ba}\right)-\gamma_{bb}\rho_{bb},     \\
& \dot{\rho}_{bc}=-i\left(\Delta_{ac}-\Delta_{ab}\right)\rho_{bc}
+\frac{i}{2}\left(\Omega^{\ast}_{\rm p}\rho_{ac}-\Omega_{\rm c}\rho_{ba}\right)-\gamma_{bc}\rho_{bc},   \\
& \dot{\rho}_{cc}={\rm Im}\left(\Omega_{\rm
c}\rho_{ca}\right)-\gamma_{cc}\rho_{cc}.
  \eqnum{A3} \label{eqset}
\end{array}
\right.
\end{equation}
Here dot denotes the derivative with respect to time. $\Omega_{\rm
p}(=\Omega_{ab})$ and $\Omega_{\rm c}(=\Omega_{ac})$ are the Rabi
frequencies of the probe and coupling light, respectively. The
frequency detunings are defined as follows:
$\Delta_{ab}=\omega_{a}-\omega_{b}-\omega_{ab}$,
$\Delta_{ac}=\omega_{a}-\omega_{c}-\omega_{ac}$ and
$\Delta_{bc}=\Delta_{ac}-\Delta_{ab}$. It should be noted that if
the probe and coupling light fields are in resonance with the $ab$
and $ac$ transitions, respectively, namely, the following
conditions are satisfied: $\Delta_{ab}=\Delta_{ac}=\Delta_{bc}=0$,
Eqs. (\ref{eqset}) can be simplified into Eq. (\ref{eq1}) that has
been considered in this paper.


\section*{Figure Caption}

Fig. 1. The schematic diagram for the $\Lambda$-type EIT system.
The level pairs $|a\rangle$-$|b\rangle$ and
$|a\rangle$-$|c\rangle$ couple the probe and coupling fields,
respectively.

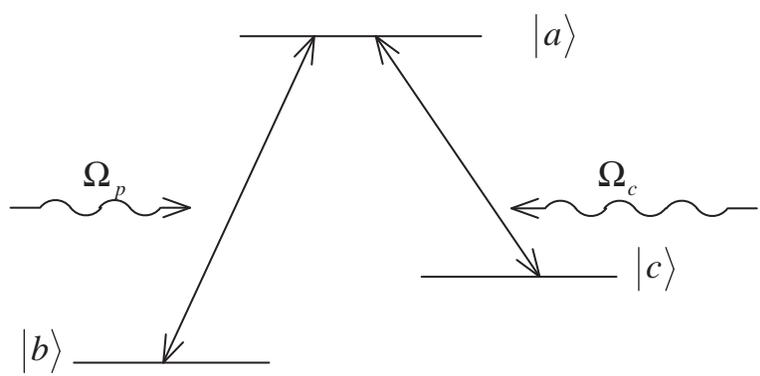
\end{document}